\documentclass[graybox]{svmult}

\usepackage{mathptmx}       
\usepackage{helvet}         
\usepackage{courier}        
\usepackage{type1cm}        
\usepackage{makeidx}         
\usepackage{graphicx}        
\usepackage{multicol}        
\usepackage[bottom]{footmisc}
\usepackage{color}
\usepackage{amsmath}
\usepackage{amssymb}

\usepackage[numbers,compress]{natbib}
\usepackage{url}

\makeindex             

\begin{document}

\title*{Gene regulatory network inference: an introductory survey}
\author{V\^an Anh Huynh-Thu and Guido Sanguinetti}
\institute{ V\^an Anh Huynh-Thu \at Department of Electrical Engineering and Computer Science, University of Li\`ege, Li\`ege, Belgium\\\email{vahuynh@uliege.be}
\and Guido Sanguinetti \at School of Informatics, University of Edinburgh, Edinburgh, U.K. \\\email{gsanguin@inf.ed.ac.uk}}
%
%
\maketitle


\abstract{Gene regulatory networks are powerful abstractions of biological systems. Since the advent of high-throughput measurement technologies in biology in the late 90s, reconstructing the structure of such networks has been a central computational problem in systems biology. While the problem is certainly not solved in its entirety, considerable progress has been made in the last two decades, with mature tools now available. This chapter aims to provide an introduction to the basic concepts underpinning network inference tools, attempting a categorisation which highlights commonalities and relative strengths. While the chapter is meant to be self-contained, the material presented should provide a useful background to the later, more specialised chapters of this book.}


\section{Introduction: the biological problem}
\label{intro}
The discovery of the biochemical basis of life is one of the great scientific success stories of the past century. Remarkably, the amazing diversity of life can be explained from a relatively small set of biochemical actors and their interactions. Heritable information is stored in chromosomes, very long  polymers of double stranded DNA, which encode information as a sequence of symbols from a four letter alphabet, A,C,G,T, the nucleotides constituting the building blocks of DNA. Just as DNA is the universal information storage medium,  information flow also follows a consistent biochemical pathway across all organisms. Stored information can be dynamically read through the process of {\it gene expression}, a two step process whereby DNA gets transcribed into RNA, an intermediate, single stranded polymer of nucleic acids (with the T nucleotide replaced by uracil, U), and RNA is subsequently translated into proteins, macromolecules formed of amino-acids which carry out most cellular functions. This process is of such fundamental importance in biology to have earned the moniker of {\it central dogma of molecular biology} \cite{crick1970central}; it constitutes the universal flow of information across all living creatures (the most notable exception being reverse transcription of viral RNA).

Not all DNA within a cell codes for proteins, and not all DNA is transcribed; indeed, genes, the stretches of DNA encoding some functionality (either protein or other classes of functional RNAs), constitute a small fraction of the overall genome. One of the surprising outcomes of the major genome-sequencing projects at the turn of the millennium was the realisation of just how little DNA codes for proteins (approximately 3\% of the human genome, with similar percentages in other higher eukaryotes). Moreover, the number of genes in different organisms is relatively constant across scales of organismal complexity: the humble baker's yeast {\it Saccharomyces cerevisiae} has approximately 6.000 genes, more than a quarter the number of genes in the human genome. Apart from raising overdue questions on our anthropocentric worldview, the natural corollary of this observation is that complexity in life does not arise from a disparity in the number of available components (genes), but from the nature and dynamics of the interactions between such components.

Measuring interactions is difficult within live cells. On the other hand, measuring components' abundances (e.g. mRNA levels) is considerably easier, and technological advances within the last two decades have enabled increasingly large-scale measurements of gene expression at steadily decreasing costs. This trend has provided a powerful motivation to attempt to reconstruct {\it computationally} the interaction structures underpinning patterns of gene expression: these interactions collectively are denoted as {\it Gene Regulatory Networks} (GRNs). Reconstructing such networks has been a central effort of the interdisciplinary field of {\it Systems Biology}.

In this chapter, we provide a tutorial overview of the field, aimed at a novice computational scientist or biologist wishing to approach the subject. We first provide a brief introduction to the core biological concepts, as well as the main sources of data currently available. We then introduce the core mathematical concepts, and briefly attempt a categorization of the main methodological approaches available. This chapter is intended to be a self-contained introduction which will provide some essential background to the book; later chapters will describe more advanced concepts, and associated tools for GRN reconstruction across the breadth of their biological application.
\subsection{Mechanisms of gene regulation}
The molecular bases of the transcription process have been intensely studied over the last 60 years. Many excellent monographs are available on the subject; we refer the reader in particular to the classic books by Ptashne and collaborators \cite{ptashne2002genes,ptashne2004genetic} (see also this recent review \cite{ptashne2014chemistry} for a historical perspective). Here we give a brief intuitive description of the process, taking, as an illustrative example, the transcriptional response of the bacterium {\it Escherichia coli} in response to changes in oxygen availability (see reference \cite{bettenbrock2014towards} for a modern review of this field). Transcription is carried out by the enzyme RNA polymerase (RNAP), that slides along the DNA, opening the double strand and producing a faithful RNA copy of the gene. The rate of recruitment of RNAP at a gene can be modulated by the presence or absence of specific {\it transcription factor} (TF) proteins, which contain a DNA-binding module that enables them to recognise specific  DNA-sequence signals near the start of genes (promoter regions). The classical view of gene regulation holds that changes in cellular state are orchestrated by changes in binding by TFs. 

For example, in {\it E. coli}, oxygen withdrawal leads to dimerisation of the master regulator protein Fumarate Nitrate Reductase (FNR); FNR dimers (but not monomers) can bind specifically to DNA, and change the rate of recruitment of RNAP at the FNR target genes, thereby changing their levels of expression to enable the cell to adapt to the changed conditions. However, FNR is not the only regulator responding to changes in oxygen availability: another master regulator, the two component system ArcAB, also senses oxygen changes, albeit through a different mechanism, and changes its binding to hundreds of genes as a result. FNR and ArcAB share many targets, and through their combined action they can give rise to highly complex dynamics \cite{partridge2007transition,rolfe2011transcript}.

Two important observations can be made from the previous discussion. Firstly, the regulation of gene expression levels is enacted through the action of gene products themselves: therefore, in principle, one may hope to be able to describe the dynamics of gene expression as an {\it autonomous} system. Secondly, even in the simple case of the bacterium {\it Escherichia coli}, regulation of gene expression is a complex process, likely to involve the interactions of several molecular players.

In higher organisms, the basic components of the transcriptional regulatory machinery are remarkably similar. However, many more levels of regulatory control are present: in particular, chemical modifications of the DNA itself (in particular methylation of C nucleotides) and of the structural histone proteins, around which DNA is wound, can affect the structural properties of the DNA, and hence the local accessibility to the transcriptional machinery. Such effects, collectively known as {\it epigenetic modifications}, have strong associations with transcription \cite{alberts1994molecular,bird2002dna,karlic2010histone,benveniste2014transcription}, and are generally thought to encode processes of cellular memory associated with long-term adaptation or cell-type differentiation.

Finally, while we have primarily focussed on transcription, subsequent steps of gene expression are also tightly regulated: RNA processing, translation and RNA and protein degradation all provide additional levels at which gene expression can be controlled. Mechanisms of post-transcriptional control of gene expression are less well explored, but it is widely believed that such processes, mostly effected through proteins or RNAs binding to RNA targets, may be as prevalent as transcriptional controls \cite{hogan2008diverse,tebaldi2012widespread}{(see also Chapter 15 for perspectives on incorporating post-transcriptional regulation in GRN inference)}. Therefore, while a gene may have no effect on the expression of another gene at the RNA level, it may well be extremely important for the protein expression.

\subsection{High throughput measurements techniques}
As we have seen in the previous subsection, the control of gene expression is effected through the action of gene products themselves. Naturally, in order to discover and quantify such controls, one must then be able to simultaneously measure the levels of expression of multiple genes. Measurements of gene expression have progressed dramatically in the last twenty years, with technological advances driving a seemingly unstoppable expansion in the scope of such experiments.

Proteins are the final product of the process of gene expression. Methods based on quantitative mass spectrometry have been highly effective in quantifying hundreds to thousands of proteins within samples. Despite that, intrinsic limits to their sensitivity and a relatively complex analysis pipeline mean that such methods do not yet reach the comprehensiveness of transcriptomic measurements \cite{bantscheff2012quantitative}.

Methods for assaying RNA levels have progressed immensely in the last two decades. Microarray technology first provided enormous impetus to the field in the late 90s \cite{brown1999exploring}. Microarrays consist of thousands of short fragments of DNA (probes) arranged on a substrate chip (usually glass or some other inert material); by designing probes to complement thousands of genomic regions from target organisms, one can obtain a readout of the (steady state) concentration of thousands of transcripts within a population of cells.

Microarrays represented a turning point in our ability to comprehensively measure genetic materials; however, the design of the probes implicitly defines what can be measured, biasing the assay and limiting the scope for discovery of unexpected biological facts, e.g. previously unobserved transcripts. Next generation sequencing (NGS) technologies proved revolutionary in this context. NGS provides a massively parallel implementation of DNA sequencing protocols, which enabled it to dramatically reduce costs and expand throughput. RNA-seq is the main NGS technology used to measure transcript abundances \cite{wang2009rna}: RNA from a population of cells is reverse transcribed (usually after a selective enrichment process to filter out highly abundant ribosomal RNAs), fragmented and the resulting complementary DNA is sequenced and mapped to a reference genome. The number of fragments mapped to a particular gene, suitably normalised \cite{evans2017selecting}, then gives a raw measurement of gene expression. 

One of the major success stories of NGS technologies is the ability of combining them with a variety of biochemical assays, greatly expanding the scope of potential measurements. Of particular relevance for GRNs is the ability to select fragments of DNA bound to specific proteins by a process called immuno-precipitation. Genomic material is fragmented, and an antibody specific to a particular DNA binding protein is added, allowing separation by centrifugation of the protein. The bound DNA fragments are then released, sequenced and mapped to a reference genome to identify where the protein was bound. This technique, Chromatin Immuno-Precipitation followed by sequencing (ChIP-seq), has been instrumental in obtaining {\it in vivo} mappings of possible regulatory relationships \cite{park2009chip}.

\section{Introduction: the mathematical formulation}\label{graph-intro}
In the previous section, we have given a condensed tour of the fundamental biological problem addressed in this book. We have seen that interactions between gene products are the fundamental processes underpinning the cell's ability to modulate gene expression. High-throughput measuring techniques paved the way to the use of computational statistics techniques to reconstruct statistically such interactions, a process sometimes called {\it reverse engineering}. In this section we introduce some of the fundamental mathematical concepts common to all methods for reverse engineering GRNs, see e.g. \cite{west2001introduction} for a more comprehensive review of these concepts.

\begin{figure}
\begin{center}\includegraphics[width=0.95\textwidth]{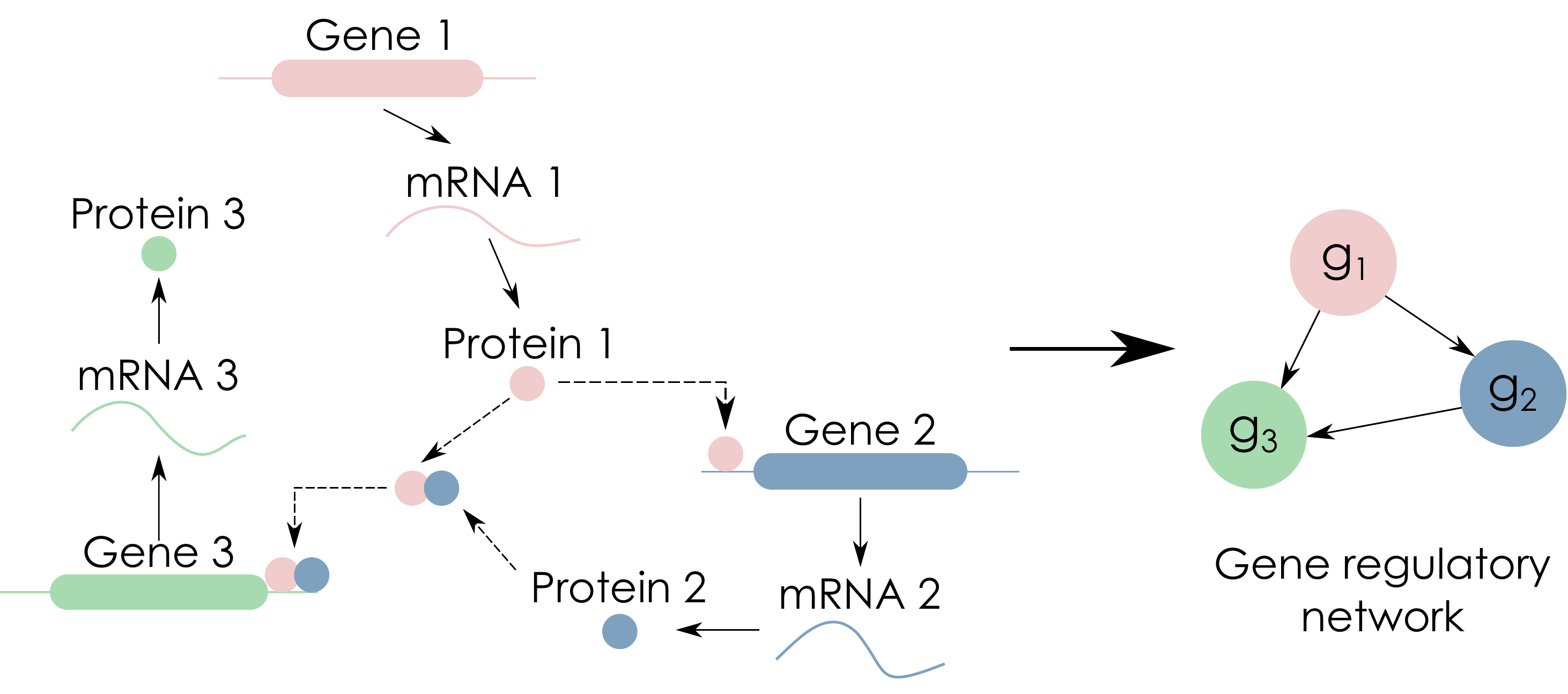}\end{center}
\caption{A cartoon schematic of a gene regulatory network. A complex biophysical model describes the interaction between three genes, involving both direct regulation (gene 2 by gene 1) and combinatorial regulation via complex formation (gene 3 by genes 1 and 2). The abstracted structure of the system is given in the (directed) network on the right.}\label{abstraction}
\end{figure}

\begin{definition}[Network] A (directed) network or graph is a pair $(V,E)$ where $V$ is a finite set of {\it vertices} (or nodes) and $E$ is a set of {\it edges} (or arcs) connecting the vertices. If $\mathcal{I}$ is a set indexing the nodes, the set of edges is a subset of the Cartesian product $E\subset\mathcal{I}\times\mathcal{I}$, with element $(ij)$ indicating the presence of an edge between node $i$ and node $j$. An {\it undirected} network is a network where  the edge set is symmetric under swapping the indices of the nodes, i.e. whenever edge $(ij)$ exists also edge $(ji)$ exists. 
\end{definition}
Within the GRN context, network nodes universally represent the expression level of genes. Edges are intuitively linked to associations between genes, but the precise meaning of an edge depends on the mathematical model of the system.
Networks are abstract representations of systems, and {\it per se} do not have a semantic interpretation that could link the network to node behaviours, e.g. their collective dynamics. Nevertheless, the structure of a network (the {\it topology}) can provide an intuitively appealing visualisation of the system, and often be informative in itself. Informally, the aim of a network abstraction is to condense in a simple representation the complexity of interactions underpinning gene expression, see Figure \ref{abstraction} for a cartoon representation. One of the most important quantities in this regard is the {\it degree} of a node, i.e. the number of edges that are attached to the node, and the {\it degree distribution} of the network, i.e. the empirical distribution of degrees across all nodes in the network. Degree distributions often encode intuitively interpretable properties of networks such as the presence of hubs or the ability to reach rapidly any node from any starting node, and in many cases they can be related to distinct stochastic mechanisms by which the network may arise. In the case of directed networks, one may further distinguish between {\it in-degree} (also called {\it fan-in}), the number of edges terminating at a node, and {\it out-degree} (also called {\it fan-out}), the number of edges starting at a node.

\begin{figure}
\begin{center}\includegraphics[width=0.95\textwidth]{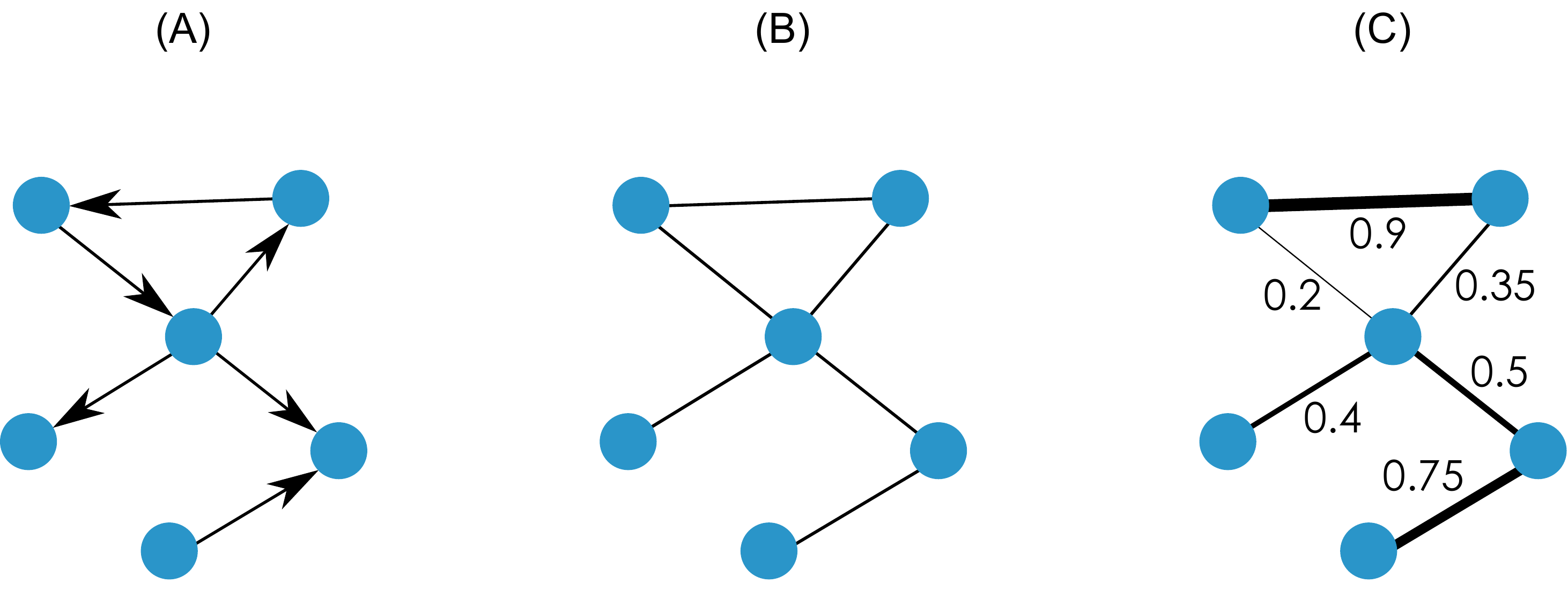}\end{center}
\caption{Examples of network types: directed (left), undirected (centre) and weighted (right), where the weights are represented by edge thickness. Note that a weighted network can be directed or undirected.}\label{netTypes}
\end{figure}

Finally, in many cases the bare topological description is insufficient in capturing aspects of interest, such as the different importances of different edges. To obviate this problem, one can consider {\it weighted} networks, where each edge is associated with a real number, its weight. We will see that in most cases reconstructed networks, the topic of this book, arise naturally as weighted networks, where the weight is intuitively associated with the support that the data offers for the existence of an edge. Weighted networks are often visualised as networks with edges of different thickness, retaining the visual immediacy of the network abstraction but effectively conveying more information. 
A schematic example of a standard graphical representation for directed, undirected and weighted networks is given in Figure \ref{netTypes}. 

Network science is a rich interdisciplinary field of research, and this whistle-stop tour of the basic mathematical concepts cannot do justice to such a field. Nevertheless, we now have the essential tools to understand, at least at a high level, many of the common strategies for reconstructing GRNs.

\section{Data-driven methods}\label{score}
The first class of GRN reconstruction methods considers a fully connected network and associates a weight to each edge by estimating gene dependencies directly from the data. The output of such methods is therefore a weighted network, which can be suitably thresholded to yield the topology of the network. Such methods are generally simple to implement, computationally efficient (they scale with the number of possible edges, which is quadratic in the number of nodes) and have proved often remarkably accurate and effective. For these reasons, some of the most popular tools for GRN inference pertain to this category. 
\subsection{Correlation networks}
The simplest score that one may associate to a pair of vector-valued measurements is their correlation. This is computed in the following way: given two zero-mean vectors $\mathbf{v}_i$ and $\mathbf{v}_j$, the (Pearson) correlation between the vectors is given by
\begin{equation}
\mathrm{corr}(\mathbf{v}_i,\mathbf{v}_j)=\rho_{ij}=\frac{\mathbf{v}_i\cdot\mathbf{v}_j}{\Vert\mathbf{v}_i\Vert\Vert\mathbf{v}_j\Vert}\label{Pearson}
\end{equation}
where $\cdot$ indicates the scalar product and $\Vert\mathbf{v}_i \Vert$ is the Euclidean norm of vector $\mathbf{v}_i$ (square root of the sum of the squares of the elements). Practically, given a set of $N$ expression measurements (e.g., different conditions) for $G$ genes, one arranges them into a data matrix $D\in\mathbb{R}^{N\times G}$. Computing correlations between the columns of $D$ yields a $G\times G$ matrix of pairwise gene correlations, which can be taken as the weights of an undirected network and suitably thresholded to obtain a network structure. Variations of this approach involve taking a different measure of correlation (e.g. Kendall's or Spearman's correlation), or raising each correlation to a power to effectively filter out spurious low correlations (weighted correlations).

Correlation networks are extremely simple to implement; their complexity scales linearly with the number of experiments and quadratically with the number of genes, so they can be easily deployed on genome-wide studies with very high numbers of experiments. The assumption that interacting genes should have correlated expression is biologically plausible, and methods such as WGCNA (weighted gene coexpression network analysis \cite{zhang2005general}) have proved consistently reliable and are widely adopted. 

Correlation networks however also have some limitations. First, two genes might appear correlated not because they genuinely interact, but because of the effect of a third gene (or several other genes). For example, a high correlation might appear between two genes that share a common regulator. Correlation networks are also unable to distinguish between direct and indirect interactions: if gene $i$ regulates gene $j$ which in turn regulates gene $k$, it is likely that there will be a high correlation between gene $i$ and gene $k$. Correlation networks are therefore vulnerable to false positives. In this respect, {\it partial correlation networks} (see Section \ref{sec:GGM}) offer a conceptually appealing solution to the problem, at the cost of some additional assumptions. Another drawback of correlation networks is that limited sample sizes (which are common in small to medium scale studies) may produce apparent high correlations which are not statistically significant. Furthermore, Pearson correlation is a linear measure of correlation, therefore non-linear regulatory effects might easily be missed, creating a vulnerability to false negatives as well.

Since the correlation is a symmetric metric, correlation networks are intrinsically undirected. Also, correlation is purely a measure of statistical association; therefore, these models are not predictive, in the sense that knowledge of some node values would not allow us to make a quantitative prediction about the remaining nodes.
\subsection{Information theoretic scores}
As we have seen before, the linearity of Pearson correlation may limit its suitability to capture complex regulatory relations. To obviate this problem, several groups have considered alternative scores based on information theory. The main mathematical concept is the mutual information, defined as follows. Let $X$ and $Y$ be two discrete random variables, and let $P(X,Y)$ be their joint probability distribution. The mutual information between the two random variables is then defined as\begin{equation}
\mathrm{MI}[X,Y]=\sum_{x_i,y_j}P(x_i,y_j)\log\frac{P(x_i,y_j)}{P(x_i)P(y_j)}=\sum_{x_i,y_j}P(x_i,y_j)\log\frac{P(x_i\vert y_j)}{P(x_i)}\label{MutInf}
\end{equation}
where $x_j$ and $y_j$ are the values the two random variables can take, and $P(X)$ (resp. $P(Y)$) is the marginal distribution obtained by summing out the values of $Y$ (resp. $X$) in the joint distribution. Intuitively, the Mutual Information quantifies the degree of dependence of the two random variables: it is zero when the two random variables are independent (as is clear from the second formulation in equation \eqref{MutInf}), and, when the two variables are deterministically linked, it returns the entropy of the marginal distribution. The mutual information is still a symmetric score, so mutual information networks are naturally undirected. Nevertheless, it can accommodate more subtle dependencies than the linear correlation score in \eqref{Pearson}, therefore potentially catering for a broader class of regulatory interactions.

In the GRN context, the idea is to replace the probability distributions in \eqref{MutInf} with empirical distributions (estimated from the samples) of gene expression levels for each pair of genes. This gives a weight to each possible edge within a fully connected, weighted undirected network; thresholding at a user-defined parameter then returns a network topology {called {\it relevance network}} \cite{butte1999mutual}. A number of methods have been proposed to filter out indirect or spurious links in relevance networks, the most popular methods being ARACNE \cite{margolin2006aracne}, CLR \cite{faith2007large} and MRNET \cite{meyer2007information}.

Mutual information networks are among the most widely used GRN inference methods. They scale to genome-wide networks, even if they are slightly more computationally intensive than correlation-based methods, as their computational complexity is quadratic in the number of genes and samples. However, they also stop short of providing a predictive framework. Furthermore, estimation of the joint probabilities in equation \eqref{MutInf} might be highly sensitive to noise when the sample size is medium-small.

\subsection{Regression-based methods}
An alternative approach to quantify the dependence of two variables is to predict one from the other. In the simplest case, one may try a linear regression approach, where the slope of the regression line may be used to quantify the strength of the dependence. In a GRN context this would amount to regressing each gene in turn against all other genes in order to obtain network weights. Thus, for every gene $g$, denoting by $x_{gi}$ its expression level in sample $i$, we would solve the regression problem\begin{equation}
x_{gi}=\sum_{j\neq g}w_jx_{ji}+\epsilon_i,\label{LinReg}
\end{equation}
with $\epsilon_i$ a noise term, and use the resulting weight $w_j$ as the weight associated with the network edge between gene $j$ and gene $g$. Notice that in this case the regression formulation naturally gives a direction to the network (even though bidirectional edges are clearly possible).

This idea is at the core of several successful methods for GRN reconstruction. TIGRESS \cite{haury2012tigress} adopts directly the framework of equation \eqref{LinReg}, introducing a L1 regularisation penalty{, which forces some of the weights $w_j$ to be strictly zero,} to ensure the identifiability of the system (in general, unless the number of samples is higher than the number of genes, these are overparametrised systems). An alternative idea is to replace the linear regression model of \eqref{LinReg} with a more flexible, non-parametric regression model. GENIE3 \cite{huynhthu2010inferring}, another widely used method, and subsequent developments \cite{van2014gene,huynh2015combining} also follow this strategy, replacing linear regression with an ensemble of {regression} trees. The score for the edge $(jg)$ is then the importance of gene $j$ in the tree model predicting gene $g$, which can be interpreted as the fraction of variance of the expression of gene $g$ that can be explained by gene $j$. Finally, the regression approach is also extremely popular to handle time series data, with the simple modification that the expression of gene $g$ at time $t$ is regressed against the expression of the other genes {\it at the previous time point} $t-1$ ({\it autoregressive model}) \cite{michailidis2013autoregressive}.{In this book, regression-based methods are discussed in Chapters 8 and 9.}

Methods based on a regression approach are amongst the most popular and scalable approaches for reconstructing directed networks. Compared to other data-driven methods, they are generally computationally more intensive, but they have predictive capability, in the sense that, given the expression of a subset of genes, one may in principle predict the expression levels of the remaining genes. {Moreover, regression-based methods are potentially able to capture high-order conditional dependencies between gene expression patterns, while correlation- and mutual information-based methods only focus on pairwise dependencies.} Practically, the identifiability of {regression} models from limited data may be problematic: different genes often have strongly correlated expression patterns, and (regularised) regression with correlated covariates is notoriously prone to spurious results. 
\section{Probabilistic models}
The data-driven based methods described before all start from some statistical or information theoretic measure of dependence, but do not explicitly formulate a model of the data in terms of probabilities. In this section, we briefly introduce two classes of methods that start explicitly from a probabilistic model of the data, using global measures of fit (joint likelihood) or Bayesian approaches to identify the network structure. 
\subsection{Gaussian Graphical Models}\label{sec:GGM}
The simplest probabilistic model one may wish to consider is a multivariate normal distribution. The probability density for a multivariate normal vector $\mathbf{x}\in\mathbb{R}^G$ is given by\begin{equation}
p(\mathbf{x}\vert\mathbf{m},\Sigma)=\frac{1}{\sqrt{2\pi\vert\Sigma\vert}}\exp\left[-\frac{1}{2}\left(\mathbf{x}-\mathbf{m}\right)^T\Sigma^{-1}\left(\mathbf{x}-\mathbf{m}\right)\right]\label{MVNdensity}
\end{equation}
where the mean vector $\mathbf{m}$ and variance-covariance matrix $\Sigma$ represent the parameters of the distribution. The off-diagonal entries of the symmetric matrix $\Sigma$ give the covariance between different entries of the random vector $\mathbf{x}$, which is related to the correlation via multiplication by the marginal standard deviations.

An important result is that the inverse of the variance-covariance matrix, the {\it precision matrix} $C=\Sigma^{-1}$, contains the {\it partial correlations} between entries in the random vector $\mathbf{x}$. The partial correlation represents the residual correlation between two variables once the effect of {all the other} variables has been removed. As such, it provides a better measure of association than simple correlation, as it is less vulnerable to spurious associations.

This insight has been effectively used in the context of GRNs by a class of models known as {\it Gaussian Graphical Models} \cite{schafer2004empirical}. The idea is to treat gene expression measurements as a multivariate normal random vector (each entry of the vector representing the expression of one gene), and then estimate the precision matrix from multiple conditions using maximum likelihood estimation. Since this requires estimating a number of parameters which is proportional to the square of the number of genes, regularisation techniques are needed; sparse regularisation techniques such as L1 regularisation (also known as {\it graphical lasso} \cite{graphlasso}) have the added advantage of returning a more interpretable result, with the non-zero entries of the precision matrix representing the edges of the (undirected) regulatory network. Several algorithmic approaches have been proposed to carry out this estimation efficiently, and Gaussian Graphical Models represent a popular network inference approach.{Within this book, Chapter 6 discusses the most recent developments in Gaussian Graphical Models usage.}

While Gaussian Graphical Models are certainly a success story, as usual they come with limitations. Estimating a high-dimensional precision matrix from limited data is difficult, and, while using a consistent estimator such as penalised maximum likelihood brings guarantees in the infinite sample limit, the accuracy of the reconstruction for finite samples is more difficult to quantify a priori. More problematically, Gaussian Graphical Models assume normality of the data, which implies linearity in the relationship between the various genes. While this can be a reasonable approximation, and surprisingly effective inferentially, it certainly is a strong modelling limitation.
\subsection{Bayesian Networks}
All methods described so far address the problem of network reconstruction from a top-down approach: start with a fully connected network, compute pairwise scores (or estimate jointly a precision matrix in the case of Gaussian Graphical Models), and then threshold/ regularise to obtain a sparse network structure. In this subsection we will briefly introduce a very popular class of methods that takes the opposite approach, constructing a joint probabilistic model out of local conditional terms, {\it Bayesian networks}.

\begin{figure}[t]
\sidecaption[t]
\includegraphics[width=3cm]{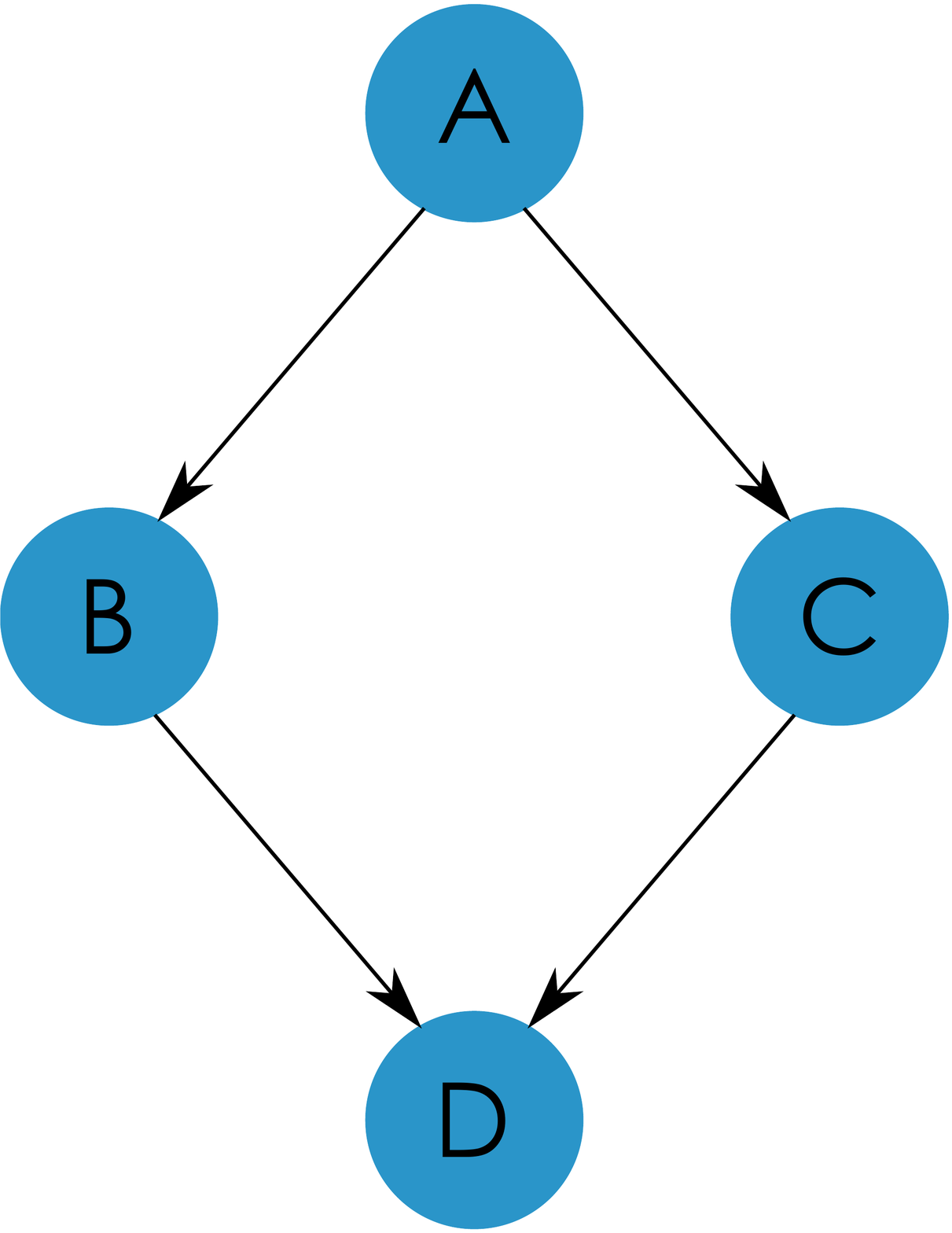}
\caption{Example of a valid Bayesian Network with four nodes and four edges. Given this structure $\mathcal{G}$, the joint distribution $P(A,B,C,D|\mathcal{G})$ factorises as $P(A)P(B\vert A)P(C \vert A)P(D \vert B,C)$.}
\label{BayesNet}      
\end{figure}

The starting point is the product rule of probability, which holds that for any two random variables $X$ and $Y$, $P(X,Y)=P(X\vert Y)P(Y)$. Applying this rule recursively, one has that for $G$ variables\begin{equation}
P(X_1,\ldots,X_G) = P(X_1) \prod_{i=2}^G P(X_i \vert X_1, \ldots, X_{i-1})
\label{prodRule}
\end{equation}	
This factorisation is general and clearly not unique, since the ordering of the random variables is arbitrary. Bayesian networks start from this general factorisation, but create structure by imposing that only a subset of all possible variables are relevant in the conditioning set \cite{pearl2014probabilistic}. More formally, for each variable $X_i$, we define the set of {\it parents} of $X_i$, $\pi_i\subset\{X_1,\ldots,X_{i-1},X_{i+1},\ldots,X_G\}$. We then construct a directed network by connecting parents and children (the direction of the arrow goes from parents to children); the network structure corresponds to a special factorisation of the joint probability as \begin{equation}
P(X_1,\ldots,X_G\vert\mathcal{G})=\prod_{i=1}^G P(X_i\vert X_{\pi_i})
\end{equation}
where we introduce the variable $\mathcal{G}$ to denote the graph structure of the Bayesian network. When the parent set $\pi_i$ is empty, the conditional distribution $P(X_i\vert X_{\pi_i})$ is equal to the marginal distribution $P(X_i)$. See Figure \ref{BayesNet} for an example. Two remarks are important: not all parents-children assignments will lead to a valid factorisation of the joint probability distribution. A fundamental result is that only networks without directed loops ({\it directed acyclic graphs, DAGs}) specify valid probability distributions (i.e. you cannot return to the same place walking on the network along the direction of the arrows). This global constraint poses considerable difficulties to reconstruction algorithms. Furthermore, even with the DAG constraint, the correspondence between networks and probability distributions is not one-to-one. As already highlighted in the case of the factorisation \eqref{prodRule}, there can be multiple valid factorisations of a joint probability distribution, leading to different networks encoding exactly the same probability distribution. This issue is known as {\it Markov equivalence} in probability theory; see e.g. \cite{barber2012bayesian}  Ch. 3 for more details about the mathematical  aspects of graphical statistics.

Within a GRN context, Bayesian networks have been hugely popular due to the simplicity with which prior information (e.g. in the form of known interactions) can be incorporated{(see for example Chapter 7 for applications of this paradigm to modern problems)}. As usual, gene expression levels are taken to represent the nodes of the network. For computational convenience, all conditional distributions are generally assumed to be Gaussian or discrete (multinomial), which enables the distributional parameters to be efficiently marginalised. In this way, one can easily compute the marginal likelihood function by evaluating the probability of the data under the model. The outstanding problem then remains the identification of the network structure. This is a very difficult combinatorial optimisation problem. Greedily searching the space of networks structures for an optimum of the likelihood was an early solution \cite{friedman2000using}: although this can be surprisingly effective, 
 in practice the cardinality of the space of network structures increases super-exponentially with the number of nodes, creating a formidable computational problem. This problem is compounded by the existence of multiple optima (due to Markov equivalence) and by the fact that the search must be constrained by the global DAG condition. As an alternative, Bayesian statistical methods have been extensively studied. This approach usually proceeds by constructing a biased random walk in the space of allowable network structures such that structures with a higher posterior probability are visited more often (a procedure called Markov Chain Monte Carlo) \cite{friedman2003being}. The Bayesian approach has considerable advantages in the ease with which prior information can be encoded, and in the way the intrinsic uncertainty in the system is represented: typically, such methods return an ensemble of plausible network structures, weighted by their posterior probability. Nevertheless, Bayesian methods suffer from considerable computational overheads and, despite recent advances \cite{hill2012bayesian}, the scalability of Bayesian network methods to genome-wide data sets remains a challenge. 
\section{Dynamical models}
One of the central questions in biology is  how organisms adapt to changing conditions. Therefore, a substantial fraction of high-throughput experiments have a time series design, e.g. they assay the same system at different time points to follow the evolution of the system in time. GRNs play a fundamental role in the mathematical modelling of such processes; unsurprisingly, several GRN reconstruction techniques are tailored towards the analysis of time series data. In this section, we introduce two broadly used classes of methods to infer network structures from dynamic data.

\subsection{Dynamic Bayesian Networks}
As we have seen in the previous chapter, a fundamental requirement on the structure of a Bayesian network is the absence of loops (DAG condition). Within the GRN context, this has long been seen as one of their main limitations: biological systems often exhibit feedback loops as a mechanism to engender robustness and stability. An elegant solution is provided by {\it Dynamic Bayesian Networks} (DBNs), a special class of Bayesian networks adapted for time series data. 

\begin{figure}
\begin{center}\includegraphics[width=0.75\textwidth]{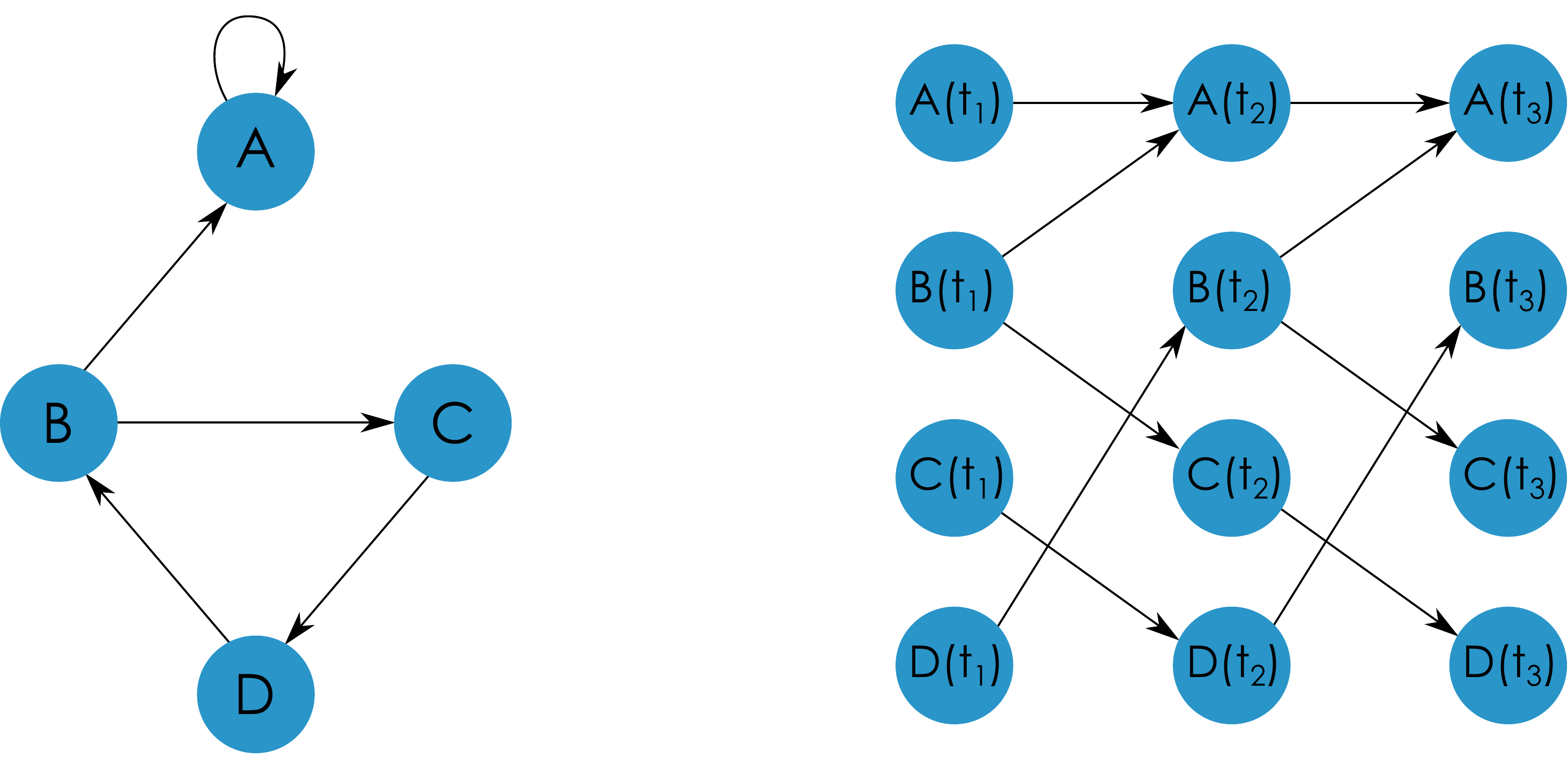}\end{center}
\caption{Example of a Dynamic Bayesian Network with four nodes: static representation (with cycles) on the left, and unrolled dynamic representation on the right.}\label{DBN}
\end{figure}

DBNs work around the DAG condition by expanding the set of random variables under consideration, so that the nodes of the network now represent expression of genes {\it at a specific time point}. Network edges may now only connect nodes pertaining to different time points, so that a gene can only influence the expression of another gene (or, indeed, itself) at a later time point ({see Figure \ref{DBN} for an example}). In this way, the DAG condition is automatically satisfied, while at the same time biologically plausible features such as feedback mechanisms can be easily incorporated. In most cases, the dynamic structure of a DBN is chosen such that edges are only present between nodes at consecutive time points, with time-independent transition probabilities: this assumption of a homogeneous, first order Markov process is a plausible approximation in many cases, and, particularly when the conditional distributions are chosen to be Gaussian, it allows the modeller to leverage a rich literature on signal processing in autoregressive models.

DBNs are extremely popular in the GRN context, and are implemented in several software tools (see \cite{oates2012network} for a recent review,{and also Chapters 2 and 3 in the present volume}). Structure learning within DBNs is easier than in standard Bayesian Networks, since the DAG condition is automatically satisfied, however it still remains computationally demanding, particularly in a Bayesian setting. From the modelling point of view, most implementations assume a linear dynamic model, which is clearly a limitation. Extensions exist which include nonlinear mappings between time points \cite{morrissey2010reverse,aijo2009learning} or that relax the time-homogeneity assumption \cite{grzegorczyk2011non}, however these incur generally higher computational costs and/or place strong restrictions on the class of nonlinear functions allowed. Most often, DBNs are implemented so that each time point in the model corresponds to an observation time. While this is somewhat natural, it constrains all biological processes to have essentially the same time-scale, which can be a serious limit; this is addressed by using a continuous-time semantic within the model, as in the case of continuous time Bayesian Networks \cite{nodelman2002continuous} or, more generally, of differential equation models.
\subsection{Differential equation methods}
Differential equations represent perhaps the best studied and most widely used class of dynamical models in science and engineering. They provide an infinitesimal description of the system dynamics by relating the rate of change (time derivative) of a variable to its value, \begin{equation}
\frac{d\mathbf{x}}{dt}=\mathbf{f}\left(\mathbf{x},\Theta,\mathbf{u}(t),t\right).\label{ODE}
\end{equation}
Here $\mathbf{f}$ is a general, time dependent, vector valued function of the variable of interest $\mathbf{x}$ itself, taking as additional inputs a set of parameters $\Theta$ and possibly also a set of external signals $\mathbf{u}(t)$. When the function $\mathbf{f}$ does not depend explicitly on time, the system is said to be {\it time homogeneous}, and when it does not depend on external inputs it is said to be {\it autonomous}.

Within a GRN context, the variables $\mathbf{x}$ are the expression levels of the set of genes we are interested in modelling, and the interactions between genes are encoded in the parameters $\Theta$. By far the most widely used class of models are linear, autonomous and time homogeneous models, where equation \eqref{ODE} simplifies to\begin{equation}
\frac{d\mathbf{x}}{dt}=A\mathbf{x}\label{linearODE}
\end{equation}
where the parameters $\Theta$ form the interaction matrix $A$. A non-zero entry $A_{ij}$ signifies an influence of gene $j$ on the time evolution of gene $i$, and hence a directed edge between $j$ and $i$ in the GRN.

Equation \eqref{linearODE} or variants thereof are at the core of several methods for inferring GRNs. The Inferelator \cite{bonneau2006inferelator} is one such popular approach, where the derivative on the left hand side of \eqref{linearODE} is approximated with the difference of observed values at consecutive time points, and the network structure is recovered via L1 regularised regression. Other approaches solve directly the differential equation \eqref{linearODE}, positing the solution to be a linear combination of basis functions \cite{trejo2015bayesian} or a draw from a Gaussian process \cite{dondelinger2013ode}, and then take a Bayesian approach to infer the parameters of the differential equation under a suitable, sparsity inducing prior distribution. Finally, the restriction to linear dynamics is not central to methods based on differential equations, and indeed methods using non-linear dynamics (such as Hill kinetics \cite{mcgoff2016local}) have been proposed.{See Chapter 16 for a comprehensive description of state-of-the-art methods for inferring GRNs using differential equations}.

Differential equation models offer several potential advantages: their continuous-time semantics is closer to the class of models used in biophysical approaches to systems biology \cite{alon2006introduction}, so that in principle such approaches can benefit from a more mechanistic interpretation. Employing a continuous-time semantics also has the added advantage of limiting the influence of experimental design decisions (e.g. choice of time points/sampling frequencies) on the final result. In other respects, differential equation models are subject to the same computational hurdles as other methods, and they suffer from similar identifiability issues. 
\section{Multi-network models}
All of our previous discussion has assumed that all the data can be explained by a single network structure. While this may be reasonable when all data comes from similar conditions, it is a very strong assumption when one is trying to jointly model data from heterogenous scenarios, as different biological conditions may lead to different pathways being activated, so that effectively different network structures may be more appropriate.

This idea has been fruitfully exploited in two main directions. Several papers have considered the scenario where data (e.g. time series) is available from different, but related conditions. Therefore, one may reasonably assume some commonalities between the underlying network structures, so that methods that can transfer information across conditions are needed. This transfer can be achieved via introducing a shared diversity penalty within different optimisation problems \cite{niculescu2007inductive,Chiquet2011}. Equivalently but more flexibly, {the joint reconstruction of the different networks can be achieved} by adopting a hierarchical Bayesian approach \cite{werhli2007reconstructing,penfold2012nonparametric}.

Another direction that has seen considerable interest is the idea of {\it time-varying networks}. Here, the assumption is that the network structure itself can rewire across time, for example to account for checkpoints during development or cancer evolution. The solution is generally composed of two steps: the identification of the change-points, and a joint learning of related networks across the homogeneous stretches of the time series. This idea has been explored both in the context of optimisation approaches \cite{ahmed2009recovering,robinson2010learning}, and  more extensively in a Bayesian scenario \cite{lebre2010statistical,thorne2012inference,dondelinger2013non}. 

Some of these ideas are explored in Chapters 2, 10 11 and 13 of this volume.


\section{Evaluation}
During our discussion of various methods for GRN inference, we have often referred to several methods as successful or effective, without specifying how the performance of a particular method may be evaluated. This is a difficult issue: GRN inference methods are motivated precisely by the difficulty of directly measuring regulatory relationships between genes, therefore almost by definition gold standard scenarios where such interactions are known with high confidence are rare. One possibility is the recourse to simulated data. One may employ a biochemically plausible interaction model to generate some simulated gene expression measurements, and then evaluate the accuracy of the method against this gold standard. This strategy has been advocated by major international initiatives such as the Dialogue for Reverse Engineering of Models (DREAM) \cite{marbach2012wisdom}, which has organised a long-running challenge on GRN inference, providing both a stimulus and a benchmark for methodological development. Another direction has been the use of synthetic gene circuits as a benchmark for network reconstruction algorithms. The most well known example of this is probably the IRMA network \cite{cantone2009yeast}, a synthetic network of 5 genes engineered within living yeast cells. While this synthetic biology approach is appealingly close to biological reality, so far technological limitations mean that such an approach has been limited to small networks containing a handful of genes.

Having decided on a benchmark data generation procedure, the next step in evaluating a GRN inference algorithm is the choice of a suitable metric. Naively, one may consider thresholding the algorithm's outputs and reporting an average accuracy in detecting presence or absence of edges. This strategy is however flawed since GRNs are typically very sparse, so that an algorithm constantly predicting the absence of edges would potentially achieve high accuracy. A better strategy is to consider the fraction of true positive calls relative to all positives ({\it sensitivity} or {\it recall}) and the fraction of true positive calls out of all positive calls ({\it precision} or {\it positive predictive value}).

\begin{figure}
\begin{center}\includegraphics[width=0.95\textwidth]{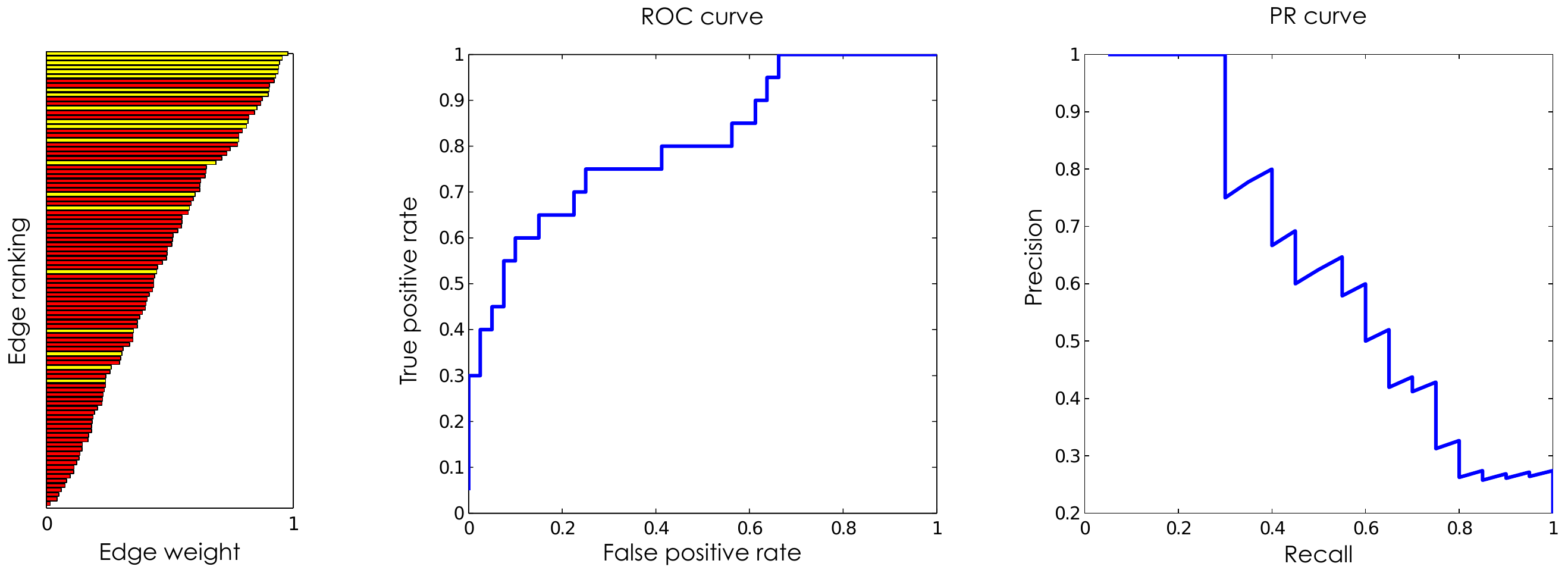}\end{center}
\caption{Evaluation of inferred networks: an algorithm typically outputs a ranked list of edges, with the weight of each edge being given by either a score or a posterior probability (left panel, where true and false edges are coloured in yellow and red, respectively). By progressively lowering the threshold for acceptance, one can construct either a ROC curve (central panel) or a precision-recall curve (right panel).}\label{ROC_PR}
\end{figure}

Naturally, precision and recall depend on the threshold chosen: with a very lax cutoff, we will likely recall many true positives (high recall), at the cost of many false positives (low precision). To elucidate the effectiveness of an algorithm in handling the precision/recall trade-off, a visually appealing strategy is the use of {\it precision-recall curves}. These are constructed as follows: given the output of a GRN inference algorithm as a weighted network, one starts by thresholding at a very strict (high) threshold, where precision {is expected to be} high and recall will be low. Decreasing the threshold, one will progressively lower precision by introducing some false positives, but also increase recall, until at zero threshold (fully connected network) recall is 1 and precision is the fraction of actual edges over possible edges (positives fraction). This procedure results in a curve in precision-recall space (see Figure \ref{ROC_PR}, right panel, for an illustration) indicative of the overall performance of the inference algorithm: a random predictor will always have an expected precision equal to the positives fraction, while an ideal algorithm will have precision 1 for any {recall between 0 and 1}. These observations justify the use of the {\it area under the curve} as a global metric of performance for an algorithm, a choice almost universally adopted in evaluating GRN inference methods.

{Alternatively, a receiver operating characteristic (ROC) curve may be used to evaluate a weighted network against a gold standard. A ROC curve plots  the recall versus the {\it false positive rate} (the fraction of false positive calls relative to all negatives) for different thresholds on the weights, again progressively lowering the threshold. Precision-recall curves are however more suited than ROC curves for problems where the number of negatives is much higher than the number of positives, which is typically the case of GRNs} \cite{davis2006relationship}.
\section{Software tools}
Most of the methods described above have been implemented in software tools which have been made freely available to the community. As it is perhaps to be expected of such a diverse and dynamic field, no single method has yet emerged as an industry standard, and tools differ widely in their usability and implementation. We provide here a summary list of some of the main software tools, as a reference list for the practitioner. All information is up-to-date at the time of writing (November 2017), and may clearly change. Naturally, this list is incomplete, and we would like to stress that any omissions do not reflect a judgement on the methods, but rather a restriction in space.
\begin{itemize}
\item{WGCNA, weighted correlation network analysis, an R package available from the comprehensive R archive CRAN.\\\small\url{https://labs.genetics.ucla.edu/horvath/CoexpressionNetwork/Rpackages/WGCNA/index.html}}
\item{ARACNe, mutual information based network inference approach. Source code in C++ available, as well as several OS-compatible versions and plugins.\\
\small\url{http://califano.c2b2.columbia.edu/aracne/}}
\item{CLR, context likelihood of relatedness, mutual information based network inference approach, originally implemented in MATLAB.\\
\small\url{http://m3d.mssm.edu/network_inference.html}}
\item{MRNET, mutual information based network inference approach. R implementation available in the Bioconductor package minet (also contains R implementations of ARACNe and CLR).\\
\small\url{https://www.bioconductor.org/packages/release/bioc/html/minet.html}}
\item{GENIE3 and other tree based methods, available as MATLAB, Python and R packages.\\
\small\url{http://www.montefiore.ulg.ac.be/~huynh-thu/software.html}}
\item{GeneNet, R package implementing Gaussian Graphical Models network inference, available from CRAN.\\\small\url{https://cran.r-project.org/web/packages/GeneNet/index.html}}
\item{CatNet, R package for (discrete) Bayesian Network structure learning, available from CRAN.\\
\small\url{https://cran.r-project.org/web/packages/catnet/index.html}}
\item{Banjo, Java package for Bayesian Networks structure learning.\\
\small\url{https://users.cs.duke.edu/~amink/software/banjo/}}
\item{G1DBN, R package for Dynamic Bayesian Network inference, available from CRAN.\\
\small\url{https://cran.r-project.org/web/packages/G1DBN/index.html}}
\item{GRENITS, Bioconductor package for Dynamic Bayesian Network inference.\\
\small\url{https://bioconductor.org/packages/release/bioc/html/GRENITS.html}}
\item{TSNI, differential equations based method, available as MATLAB package.\\
\small\url{http://dibernardo.tigem.it/softwares/time-series-network-identification-tsni}}
\item{Inferelator, differential equations based method.\\
\small\url{http://bonneaulab.bio.nyu.edu/networks.html}}
\item{netbenchmark, R package for benchmarking GRN inference methods (also contains R implementations of several methods such as ARACNe, C3NET, CLR, GeneNet and GENIE3).\\
\small\url{https://www.bioconductor.org/packages/release/bioc/html/netbenchmark.html}}
\end{itemize}

\section{Discussion and outlook}
GRN inference is a mature field of methodological research, with widespread and increasing applications in biomedical research. In this chapter, we have attempted a broad brush introduction to the field, highlighting the biological motivation and the technological advances in data collection that have underpinned its recent flourishing. We then proceeded to give a bird's eye view of the statistical principles underpinning some of the most popular methodologies for GRN inference. Our focus has been on the foundations, attempting a coarse categorisation of different methods based on their assumptions and semantics. Of course, many interesting contributions fall at the intersection of different categories, and are not well accommodated by our simplifying approach.

Naturally, it is impossible to do justice to a rich and wide research area within a short introductory review. Our aim here is to prepare the reader for more advanced concepts to be described in subsequent chapters of this book; nevertheless, we hope that this chapter will also form a worthwhile introduction for the novice to the field, and have attempted to make it as self contained as possible.

\begin{acknowledgement}
GS acknowledges support from the European Research Council under grant MLCS 306999. VAHT is a Post-doctoral Fellow of the F.R.S.-FNRS.
\end{acknowledgement}
\bibliographystyle{spbasic_unsort}
\bibliography{intro_bib}

\end{document}